# Initialization Process of a Power System Transient Simulation Scheme for Stability Studies

Sheng Lei[1,2], Student Member, IEEE, and Alexander Flueck[1], Senior Member, IEEE

[1]Department of Electrical and Computer Engineering  
Illinois Institute of Technology  
Chicago, IL, USA

[2]Mathematics and Computer Science Division  
Argonne National Laboratory  
Lemont, IL, USA

Email: slei3@hawk.iit.edu, and flueck@iit.edu

*Abstract*—The initialization process of a novel power system transient simulation scheme for stability studies is put forward, by further developing a "time-domain harmonic power-flow algorithm". The initialization process is formulated as an algebraic problem to ensure that the power system under study is in steady state and operated at a specified operating point, at the beginning of a transient simulation run. The algebraic problem is then solved efficiently by a preconditioned finite difference Newton-GMRES method. Case studies verify the validity and efficiency of the initialization process. The proposed initialization process is general-purpose and can be applied to other power system transient simulation schemes.

## I. INTRODUCTION

Transient simulation is a powerful tool for studying dynamic behavior of power systems [1]-[4]. For stability studies, a transient simulation run is typically required to start from the steady state [3], [4]. Furthermore, power flow conditions are also to be satisfied, which characterize the operating point of the system [3], [4]. The initialization process of transient simulation for stability studies has to meet these requirements.

For three-phase balanced power systems, initialization process is mature and available in the literature [3], [4]. Nevertheless that for general unbalanced power systems is much more complicated and difficult to achieve [1], [5]. Power flow computation [3], [4] and the multiphase version [6] are no longer applicable due to the presence of harmonics [7]. Time domain techniques for steady state computation have been applied to power systems [8], [9]; however the power flow conditions were not considered. How to initialize a possibly unbalanced power system is also a huge challenge encountered by the novel transient simulation scheme proposed in an earlier work of the authors [10], which is based on frequency response optimized integrators considering second order derivative and especially suitable for stability studies on general unbalanced power systems.

In fact, initialization process for unbalanced power systems may be linked to the "time-domain harmonic power-flow algorithm" proposed in [11]. Based on the sensitivity circuit analysis [12], [11] formulates a problem aiming at obtaining the steady state solution for general power systems so that the state variables are periodic regarding the nominal fundamental period while the power flow conditions are satisfied. That paper then uses Newton's method to solve the problem. The Jacobian matrix is constructed by the finite difference method at the first iteration; it is updated with Broyden's method in the later iterations. The initial guess of the unknowns is done by a flat start.

This paper further develops the work of [11] to put forward the initialization process of the novel transient simulation scheme [10]. Contributions of this paper are three-fold. First, the initial value condition is added to account for variables which cannot be periodic regarding the nominal fundamental period. Based on an alternative and straightforward derivation, the initialization process is formulated as an algebraic problem to simultaneously tackle the periodic boundary value condition, initial value condition and power flow conditions, which is to be solved via Newton's method. Second, a preconditioned finite difference generalized minimal residual (GMRES) method is introduced to solve each Newton iteration [13], [14]. The combination of Newton's method and the finite difference GMRES method is referred to as the finite difference Newton-GMRES method. Compared to constructing the Jacobian matrix by the finite difference method, the approach adopted in this paper significantly improves the computational efficiency. It is enhanced by an initial guess achieved based on three-phase power flow computation [6], [16], [17]. Third, the functionality of the novel transient simulation scheme [10] is strengthened in that it is able to start simulation runs from true steady states satisfying power flow conditions.

The remainder of this paper is organized as follow. Section II provides a brief overview on the objective of the initialization process. Section III formulates the initialization process as an algebraic problem. Section IV introduces the preconditioned finite difference Newton-GMRES method to solve the problem. Section V verifies the validity and efficiency of the proposed initialization process via case studies. Section VI concludes the paper and points out some directions for future research.

## II. OBJECTIVE OF INITIALIZATION PROCESS

To start a transient simulation run, the initial value of variables, the external inputs to and parameters of a power system under study have to be specified. Some parameters are intrinsic to the power system, such as the resistance and inductance of a transmission line. These parameters can be found in the input data for the simulation. Other parameters and the external inputs depend on the specific operating point. Examples of the dependent parameters include the resistance and inductance of a static load. Examples of the external inputs include the voltage reference of an excitation system and the load reference of a turbine-governor system. The objective of the initialization process is to give proper values to the variables, the dependent parameters and the external inputs.

As mentioned in Section I, stability studies require these quantities to be provided so that the power system is initially in steady state. In steady state, the dependent parameters and the external inputs should be constant. However their specific values have to be determined so that the power system is operated at a certain operating point, which is depicted by power flow conditions. The following section will detail how to meet these requirements in the initialization process.

## III. FORMULATION OF THE PROBLEM

### A. Periodic Boundary Value Condition

Note that the nominal fundamental period is also a period of any higher order harmonics. The DC offset, if there is any, is constant in steady state. Therefore after a nominal fundamental period, a typical quantity in the power system regains its original value. Suppose that $x$ is a typical variable in the simulation. $x$ should satisfy the periodic boundary value condition

$$x(t_0 + T) - x(t_0) = 0 \quad (1)$$

where $t_0$ is the starting time; $T$ is the nominal fundamental period. The residual function is

$$x(t_0 + T) - x(t_0) \quad (2)$$

### B. Initial Value Condition

Although most of the quantities in the power system satisfy the periodic boundary value condition, some quantities are exceptional. For example, the rotor angle of an induction machine regarding the synchronously rotating reference frame generally cannot regain its original value after a nominal fundamental period. If these quantities are considered as variables in the simulation, their initial value has to be specified, resulting in the initial value condition. Suppose that $y$ is one of these variables. (3) should be satisfied

$$y_0 - y(t_0) = 0 \quad (3)$$

where $y_0$ is the initial value of $y$. The residual function is

$$y_0 - y(t_0) \quad (4)$$

### C. Power Flow Conditions

Power flow conditions usually appear in the input data for the simulation. They specify the terminal conditions of individual devices in steady state in terms of voltage magnitude or angle, real or reactive power injection or consumption, and so on. Several related comments are made as follows.

- First, power flow conditions are considered conventionally as specified at busses or nodes in the power system; terms such as "swing bus", "PV bus", "PQ bus" are used. In fact, power flow conditions are assigned to devices. A bus or a node itself generates or consumes no power; it is the devices connected to the bus or node that generate or consume power. Furthermore, a three-phase synchronous generator and a three-phase load may be connected to the same bus; but their power flow conditions are specified separately.
- Second, power flow conditions in the existing input data are mainly prepared for power flow computation. Power flow computation is based on the phasor representation at nominal fundamental frequency, no matter it is for transmission or distribution systems, positive sequence or three-phase [3], [4], [6], [15]-[17]. Therefore in this paper, the voltage magnitude and voltage angle in power flow conditions are understood as the magnitude and angle of the voltage phasor respectively; the complex power is calculated by multiplying the voltage phasor and the conjugate of the current phasor; the real power and reactive power are respectively the real part and imaginary part of the complex power. Note that voltages and currents are given as waveforms by the novel transient simulation scheme [10]. Consequently a waveform to phasor conversion is necessary to take the power flow conditions into account.
- Third, most power flow conditions for three-phase devices are prepared for positive sequence power flow computation [3], [4], [15]. Therefore, the related voltage phasor magnitude and angle, real and reactive power are understood as the positive sequence values in this paper. As a result, a phase to sequence conversion [3], [4], [15] is performed to extract the positive sequence information so that the later calculations can be carried out.
- Fourth, the assumptions made in the second and the third terms are a compromise with the existing input data. If more detailed data are available in the future, the specific calculations will need minor modifications; but the overall initialization process will remain valid and feasible.

Power flow conditions for individual types of devices are discussed as follows.

*1) Vθ device:* At least one device in the power system has to be assigned with voltage phasor magnitude and angle at its terminal so that it can serve as an angle reference for the system. Such a type of devices is called Vθ device. (5) and (6) are to be satisfied

$$V\cos(\theta) - real(V_{device}) = 0 \tag{5}$$
$$V\sin(\theta) - imag(V_{device}) = 0 \tag{6}$$

where $V_{device}$ is the voltage phasor across the device; $V$ is the specified voltage phasor magnitude; $\theta$ is the specified voltage phasor angle. The residual functions are

$$V\cos(\theta) - real(V_{device}) \tag{7}$$
$$V\sin(\theta) - imag(V_{device}) \tag{8}$$

*2) PV device:* Some devices are assigned with real power output and the terminal voltage phasor magnitude. Such a type of devices is called PV device. The real power generation of a PV device is calculated as

$$S_{device} = V_{device} I_{device}^* \tag{9}$$
$$P_{device} = real(S_{device}) \tag{10}$$

where $S_{device}$ is the complex power generation of the device; $I_{device}$ is the current injection phasor from the device; * denotes the conjugate; $P_{device}$ is the real power generation of the device. (11) and (12) are to be satisfied

$$P - P_{device} = 0 \tag{11}$$
$$V - |V_{device}| = 0 \tag{12}$$

where $P$ is the specified real power generation. The residual functions are

$$P - P_{device} \tag{13}$$
$$V - |V_{device}| \tag{14}$$

*3) PQ device:* Some devices are assigned with real power and reactive power generation or consumption. Such a type of devices is called PQ device. Note that power generation is understood as negative power consumption. The specified power consumption can be made voltage-dependent. The well-known ZIP load [3] is an example. Specifically for a ZIP load

$$P = (k_{PS} + k_{PI}|V_{device}| + k_{PZ}|V_{device}|^2)P_0 \tag{15}$$
$$Q = (k_{QS} + k_{QI}|V_{device}| + k_{QZ}|V_{device}|^2)Q_0 \tag{16}$$

where $P$ and $Q$ are the specified real and reactive power consumption respectively; $k_{PS}$, $k_{PI}$, $k_{PZ}$, $k_{QS}$, $k_{QI}$ and $k_{QZ}$ are specified coefficients, $P_0$ and $Q_0$ are the specified nominal real and reactive power consumption respectively.

The real and reactive power consumptions are calculated as

$$S_{device} = V_{device} I_{device}^* \tag{17}$$
$$P_{device} = real(S_{device}) \tag{18}$$
$$Q_{device} = imag(S_{device}) \tag{19}$$

where $S_{device}$ is the complex power consumption of the device; $I_{device}$ is the current extraction phasor; $P_{device}$ is the real power consumption of the device; $Q_{device}$ is the reactive power consumption of the device. (20) and (21) are to be satisfied

$$P - P_{device} = 0 \tag{20}$$
$$Q - Q_{device} = 0 \tag{21}$$

The residual functions are

$$P - P_{device} \tag{22}$$
$$Q - Q_{device} \tag{23}$$

*4) Induction machine:* An induction machine is usually assigned with the real power consumption [16], [17]. The calculation follows (17), (18) and (20). The residual function is the same as (22).

### D. Waveform to Phasor Conversion

Several mature techniques may be adopted to convert a waveform into its nominal fundamental frequency phasor [18], such as curve fitting, fast Fourier transform (FFT) and digital filtering. In this paper, curve fitting [2] is adopted.

### E. Phase to Sequence Conversion

Given the three-phase phasors $X_a$, $X_b$ and $X_c$, the positive sequence phasor is calculated as [3], [4], [15]

$$X_+ = \frac{1}{3}(X_a + e^{j\frac{2}{3}\pi}X_b + e^{-j\frac{2}{3}\pi}X_c) \tag{24}$$

### F. Initialization Process as an Algebraic Equation Set

Based on the discussion in this section, the initialization process is formulated as an algebraic equation set, which is constructed by transient simulation runs of one nominal fundamental period

$$F(X) = 0 \tag{25}$$

where $F$ is a vector consisting of individual residual functions introduced in this section; $X$ is a vector of unknowns consisting of the variables, the dependent parameters and the external inputs, as discussed in Section II; 0 is a zero vector of proper dimension. In fact, $X$ is the input to a transient simulation run while $F(X)$ is the outcome. The initialization process forces the residual function vector to zero, which solves (25); details will be given in the next section.

## IV. SOLUTION TO THE PROBLEM

### A. Initial Guess

To start Newton's method, an initial guess of the vector of unknowns is needed. If the initial guess is close enough to the final solution, the convergence of Newton's method will be sped up [19].

Although three-phase power flow computation is not able to give the true steady state of an unbalanced power system [17], it does give a solution which is close [6], [16], [17]. Therefore a three-phase power flow computation is first performed in order to obtain nodal voltage phasors and branch current phasors of the power system network. Three-phase power flow computation is classical, details can be found in [6], [16] and [17]. Once a phasor is obtained, it can be readily translated into instantaneous value.

To initialize three-phase devices, the phase to sequence conversion is first performed at its terminal to extract the positive sequence information. The device is then temporarily initialized in the conventional way [3], [4] as if the external system is balanced. Note that in the conventional positive sequence transient stability (TS) simulation, single-phase equivalent is used [3], [4]; the same initialization is thus applicable to single-phase devices. Similarly the controllers are also initialized. During the initialization of devices, some heuristic simplifying assumptions are made to obtain explicit expressions for the variables to be initialized so that iteration is avoided. Such assumptions include neglecting the saturation and neglecting the limit. These assumptions are justified because the objective here is to achieve an initial guess which is relatively close to the final solution. However, the exact accuracy is not the main concern when choosing a starting point for the initialization process.

Due to the simplifying assumptions mentioned above, the resulting initial guess of the vector of unknowns may be incompatible in the sense that it may cause violation of the Kirchhoff's current law (KCL). Such an issue is taken care of by treating the first time step as a discontinuity event [10].

## B. Finite Difference Newton-GMRES Method

An implementation of the finite difference Newton-GMRES method is presented in Fig. 1. The implementation of Newton's method is based on [15]; the implementation of the GMRES method is based on [13]. The dimension of the vector of unknowns is assumed to be $m$. The evaluation of $F(X)$ is introduced in Section III. *tolerance* is a user defined tolerance for Newton's method. $c$ and $s$ are $m$-dimensional vectors. $Q$ is an $m \times (m+1)$ matrix. $H$ is an $(m+1) \times m$ matrix. The absolute tolerance for the GMRES method *abstol* is heuristically set to $0.5 \times tolerance$. The relative tolerance for the GMRES method *reltol* is user defined; 0.001 is recommended. $\varepsilon$ is a small number for the finite difference method; 0.0001 is a typical value. The upper triangular linear equation set is solved by backward substitution.

## C. Preconditioning

If the eigenvalues of the Jacobian matrix are tightly clustered on the complex plane, the GMRES method will quickly converge; if on the other hand the eigenvalues are widely scattered, the convergence of the GMRES method will be very slow, if at all [13]. One idea to accelerate the convergence of the GMRES method is to solve a modified well-behaved linear equation set and transform the solution into the solution to the original problem. Suppose at a certain Newton iteration, a linear equation set should be solved

$$Jx = r \quad (26)$$

where $J$ is the Jacobian matrix; $x$ is the vector of unknowns; $r$ is the vector of residuals. Instead of (26), the following preconditioned linear equation set is solved

$$(JM)w = r \quad (27)$$

where $M$ is an invertible matrix, which is called a right preconditioner. Hopefully $JM$ has tightly clustered eigenvalues. The solution to the original linear equation set is

$$x = Mw \quad (28)$$

Reference [14] reports a strategy to form the right preconditioner for the GMRES method. An initial preconditioner is first specified by the user, which can be the identity matrix of proper size; it is then updated via Broyden's method. The updates are performed at both Newton's iteration and the GMRES iteration, referred to as the outer updates and inner updates respectively, see the outer and inner "while" loops in Fig. 1. The same strategy is implemented in this paper. It can be activated if desired. The implementation is described as follows.

- Before entering Newton's iteration, the initial preconditioner $M$ is specified, which is the identity matrix. This step is added at the beginning of Fig. 1.

- Immediately after (A) in Fig. 1, the outer update is performed. Specifically if *iter* > 1, the result of the inner updates $M_0$ is passed to $M$; then $M$ is updated as

$$M = M + \frac{(\Delta X - M(F(X) - F(X - \Delta X)))}{\Delta X^T M(F(X) - F(X - \Delta X))} \Delta X^T M \quad (29)$$

- Immediately after (B) in Fig. 1, $M$ is passed to $M_0$.
- At (C) in Fig. 1, the equation is replaced with

$$\bar{z} = MQ(:, k) \quad (30)$$

- Immediately after (D) in Fig. 1, the inner update is performed. $M_0$ is updated as

$$M_0 = M_0 + \frac{(\varepsilon \bar{z} - M_0(F(X + \varepsilon \bar{z}) - F(X)))}{(\varepsilon \bar{z})^T M_0(F(X + \varepsilon \bar{z}) - F(X))} (\varepsilon \bar{z})^T M_0 \quad (31)$$

- At (E) in Fig. 1, the update of $X$ is calculated as (32) instead

$$\Delta X = MQ(:, 1:k)y \quad (32)$$

```
convergence = 0,  iter = 0,  X = X^(0)
while convergence = 0 and iter < maxiter do
    evaluate F(X)
    ρ = ||F(X)||_2
    if ρ < tolerance
        convergence = 1
    if convergence = 0
        iter = iter + 1                                              (A)
        initialize c, s, Q, H
        k = 0,  g = (ρ, 0, 0, …) ∈ R^(m+1),  errtol = max(abstol, reltol×ρ)
        Q(:, 1) = -F(X)/ρ                                            (B)
        while ρ > errtol and k < m do
            k = k + 1
            z = Q(:, k)                                              (C)
            evaluate F(X + εz)
            Q(:, k+1) = (F(X + εz)-F(X))/ε                           (D)
            for j = 1 : k
                H(j, k) = Q(:, k+1)^T Q(:, j)
                Q(:, k+1) = Q(:, k+1) - H(j, k)Q(:, j)
            H(k+1, k) = ||Q(:, k+1)||_2
            Q(:, k+1) = Q(:, k+1)/H(k+1, k)
            if k > 1
                Givens rotations for H(1:k-1, k):
                    for i = 1 : k-1
                        tmp1 = c(i)H(i, k) - s(i)H(i+1, k)
                        tmp2 = s(i)H(i, k) + c(i)H(i+1, k)
                        H(i, k) = tmp1,  H(i+1, k) = tmp2
            v = √(H(k, k)^2 + H(k+1, k)^2)
            c(k) = H(k, k)/v,  s(k) = -H(k+1, k)/v
            H(k, k) = c(k)H(k, k) - s(k)H(k+1, k),  H(k+1, k) = 0
            Givens rotation for g(k:k+1):
                tmp1 = c(k)g(k) - s(k)g(k+1)
                tmp2 = s(k)g(k) + c(k)g(k+1)
                g(k) = tmp1,  g(k+1) = tmp2
            ρ = |g(k+1)|
        solve the upper triangular linear equation set  H(1:k, 1:k)y = g(1:k)
        ΔX = Q(:, 1:k)y                                              (E)
        X = X + ΔX
```

Figure 1. Finite difference Newton-GMRES method.

## V. CASE STUDIES

The initialization process proposed in this paper is applied to a test system in this section. The test system is a modification of the well-known 3-generator 9-bus power system [20], as shown in Fig. 2. The original static load at Bus 5 is replaced with an induction motor. Note that the real power consumption of an induction motor is specified as the power flow condition [16], [17]. The original real power is thus assigned to the induction motor while the reactive power is relaxed. System information including bus parameters, branch parameters and power flow conditions can be found in [20]. Generator and induction motor dynamic parameters used in this paper are given in the appendix. Unbalance is introduced into the system by non-uniform allocation of static loads on individual phases. Specifically, the total static load at a specific bus is allocated as follows

$$S_A = \frac{1}{3}(1-k)S, \; S_B = \frac{1}{3}S, \; S_C = \frac{1}{3}(1+k)S \quad (33)$$

where $S$ is the total complex power load at the bus; $S_A$, $S_B$ and $S_C$ are the Phase A, B and C complex power load respectively; $k$ is an allocation factor, in this paper $k = 0.1$. Static loads are assumed to be constant impedance in transient simulation.

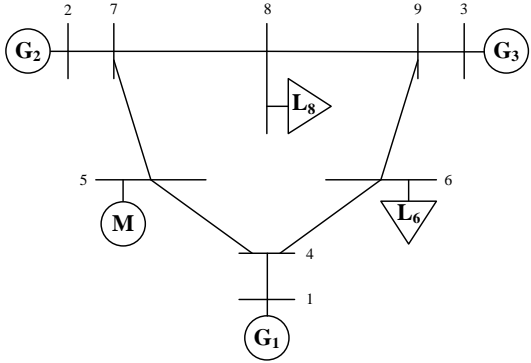

Figure 2. Modified 3-generator 9-bus power system.

For the transient simulation runs during the initialization process, a step size of one 25th of the nominal fundamental period is used (around 667 μs). The tolerance for Newton's method is set to $10^{-6}$ in Fig. 1. There are 274 elements in the vector of unknowns in (25). The convergence behavior of the initialization process is listed in Table I.

TABLE I. CONVERGENCE BEHAVIOR OF THE INITIALIZATION PROCESS

| Iteration count | 2-norm residual | |
|---|---|---|
| | With preconditioning | Without preconditioning |
| 0 | 0.3516 | 0.3516 |
| 1 | 0.0031 | 0.0031 |
| 2 | 6.7009×10⁻⁶ | 6.6987×10⁻⁶ |
| 3 | 4.3442×10⁻⁷ | 4.9758×10⁻⁷ |
| Number of transient simulation runs | 114 | 140 |

As can be seen from Table I, the finite difference Newton-GMRES method adopted in the initialization process converges to the final solution within 3 iterations, no matter the preconditioning is used or not. The preconditioning accelerates the convergence, which is reflected in the smaller number of transient simulation runs required.

If the algebraic equation set (25) is to be solved for this study case with the idea of constructing the Jacobian matrix by the finite difference method, at least 276 transient simulation runs are needed. This estimation is made even under the very generous assumption that only one Newton iteration is required, in which the first one is performed to generate the vector of residual functions, 274 are performed to construct the Jacobian matrix column by column which is equal to the number of unknowns, and the last one is performed to check the convergence criteria so that the initialization process is terminated. Nevertheless the number of transient simulation runs required by the finite difference Newton-GMRES method is much less than the number of unknowns. Therefore the finite difference Newton-GMRES method is significantly more efficient.

After the initialization process is completed, a step size of 250 μs is used for the later transient simulation. Note that such a small step size is not necessary for accuracy with the novel transient simulation scheme [10]. It is used to better demonstrate the time domain waveforms and to show that the step size used in the initialization process and that used in the later transient simulation can be different. Bus 5 Phase A voltage and Generator 2 rotor speed are plotted in Figs. 3 and 4 respectively. As the system is unbalanced, the rotor speed exhibits second order harmonics. These variables are indeed in periodic steady state, verifying the validity of the initialization process. Fig. 5 shows the induction motor rotor angle regarding the synchronously rotating reference frame. Note that it does not satisfy the periodic boundary value condition, which raises the necessity of adding the initial value condition into the initialization process.

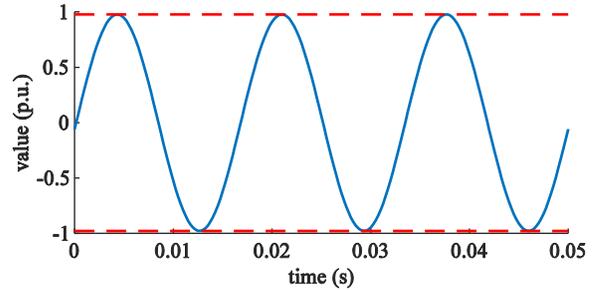

Figure 3. Bus 5 Phase A voltage.

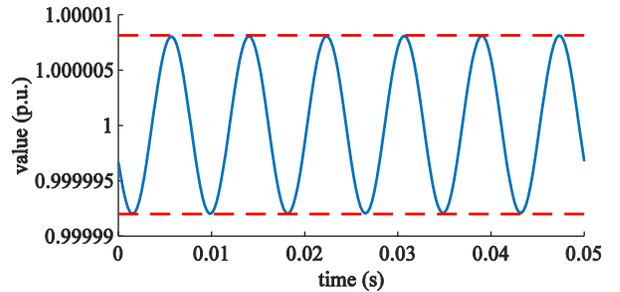

Figure 4. Generator 2 rotor speed.

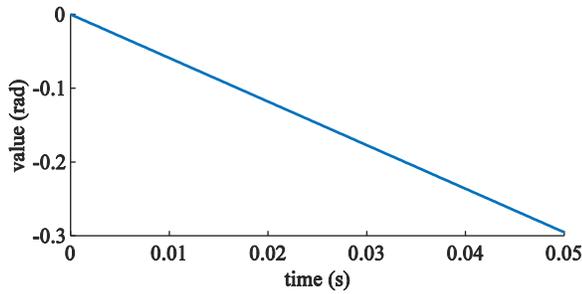

Figure 5. Induction motor rotor angle regarding the synchronously rotating reference frame.

## VI. CONCLUSION AND FUTURE WORK

The initialization process of the novel transient simulation scheme [10] is put forward in this paper. Case studies demonstrate its validity and efficiency. Although the initialization process is proposed for a specific transient simulation scheme, it can be applied to other ones. In fact, it has also been successfully implemented for an iterative electromagnetic transients (EMT) simulator based on conventional numerical integrators [10].

In the future, the initialization process may be applied to larger systems to test its scalability. Other strategies to construct the preconditioner may be investigated to further speed up the process. Research effort may be directed to constructing the Jacobian matrix analytically to improve the computational efficiency.

## APPENDIX

TABLE A. I. GENERATOR DYNAMIC PARAMETERS

| Generator (#) | 1 | 2 | 3 |
| --- | --- | --- | --- |
| Rated Power (MVA) | 100 | 100 | 100 |
| Stator Resistance (p.u.) | 0.002 | 0.002 | 0.002 |
| Stator Leakage Reactance (p.u.) | 0.0787 | 0.0787 | 0.0787 |
| D-Axis Synchronous Reactance (p.u.) | 1.575 | 1.575 | 1.575 |
| Q-Axis Synchronous Reactance (p.u.) | 1.512 | 1.512 | 1.512 |
| D-Axis Transient Reactance (p.u.) | 0.291 | 0.291 | 0.291 |
| Q-Axis Transient Reactance (p.u.) | 0.39 | 0.39 | 0.39 |
| D-Axis Subtransient Reactance (p.u.) | 0.1733 | 0.1733 | 0.1733 |
| Q-Axis Subtransient Reactance (p.u.) | 0.1733 | 0.1733 | 0.1733 |
| Open-Circuit D-Axis Transient Time Constant (s) | 6.1 | 6.1 | 6.1 |
| Open-Circuit Q-Axis Transient Time Constant (s) | 1.0 | 1.0 | 1.0 |
| Open-Circuit D-Axis Subtransient Time Constant (s) | 0.05 | 0.05 | 0.05 |
| Open-Circuit Q-Axis Subtransient Time Constant (s) | 0.15 | 0.15 | 0.15 |
| Inertia Constant (s) | 4.0 | 3.0 | 2.0 |
| Damping Coefficient (p.u.) | 0.1 | 0.1 | 0.1 |

TABLE A. II. INDUCTION MOTOR DYNAMIC PARAMETERS

| | |
| --- | --- |
| Rated Power (MVA) | 100 |
| Stator Resistance (p.u.) | 0.02 |
| Stator Leakage Reactance (p.u.) | 0.1 |
| Rotor Resistance Referred to the Stator (p.u.) | 0.01 |
| Rotor Leakage Reactance Referred to the Stator (p.u.) | 0.04 |
| Magnetizing Reactance (p.u.) | 1.75 |
| Inertia Constant (s) | 1.0 |
| Damping Coefficient (p.u.) | 0.1 |
| Stator Connection | Floating Y |